# Pressure-induced symmetry breaking in tetragonal CsAuI$_3$

Shibing Wang[1,2], Shigeto Hirai[1], Max C. Shapiro[3], Scott C. Riggs[3], Ted H. Geballe[3], Wendy L. Mao[1,4] and Ian R. Fisher[3]

[1]*Department of Geological and Environmental Sciences, Stanford University, Stanford, CA 94305, USA*

[2]*SSRL, SLAC National Accelerator Laboratory, Menlo Park, CA 94025, USA*

[3]*Department of Applied Physics, and Geballe Laboratory for Advanced Materials, Stanford University, Stanford, CA 94305, USA*

[4]*Photon Science, SLAC National Accelerator Laboratory, Menlo Park, CA 94025, USA*

**Abstract**

Results of *in situ* high pressure x-ray powder diffraction on the mixed valence compound Cs$_2$Au$^I$Au$^{III}$I$_6$ (CsAuI$_3$) are reported, for pressures up to 21 GPa in a diamond anvil cell under hydrostatic conditions. We find a reversible pressure-induced tetragonal to orthorhombic structural transition at 5.5-6 GPa, and reversible amorphization at 12-14 GPa. Two alternative structures are proposed for the high-pressure orthorhombic phase, and are discussed in the context of a possible Au valence transition.

## I. INTRODUCTION

As a fundamental thermodynamic parameter, pressure can dramatically alter materials' properties and induce structural, electronic and/or magnetic transitions. Among these, pressure-induced valence transitions have attracted wide interest because they provide a platform for studying the interplay between electronic and lattice degrees of freedom, and discovery of novel behavior. For example, in $4f$ electron systems such as rare-earth chalcogenides, the valence of the rare-earth ion can be controlled with either applied pressure or by varying the chalcogen[1]. Another interesting group of mixed valence compounds originate from strong electron-phonon coupling in certain $5d$ transition metal systems. Unlike valence fluctuations in $4f$ materials, they exhibit a mixed valence ground state with $d^{n-1}$ and $d^{n+1}$ electron configurations.

The family of alkali metal gold halides $M_2Au_2X_6$ (M = K, Rb or Cs; X=Cl, Br or I) are prototypical examples of such "valence-skipping" mixed valence $5d$ compounds, and have attracted considerable attention due to their sensitivity to applied pressure [2,3]. In particular, $CsAuX_3$ (X = Cl, Br or I) has been widely studied because of their relatively simple tetragonal perovskite structure. Evidence from early experiments indicated that pressure lead to a higher crystal symmetry through first order structural for the chloride, bromide and iodide at 12.5, 9 and 5.5 GPa respectively [4,5,6]. Additional experimental evidence based on Raman spectroscopy[7] and Mossbauer spectroscopy [8,9] suggests that coupled to the structural transition there is an associated mixed valence (MV) to single valence (SV) transition, although it is not clear whether these necessarily occur at exactly the same pressure and temperature[6]. In order to understand the origin of the suppression of this charge density wave state with applied pressure, it is necessary to carefully investigate the evolution of the crystal structure, which is the purpose of the current work.

Among the alkali metal gold halides, the critical pressure for the structural transition is lowest for $CsAuI_3$, presumably due to the increased hybridization between Au $5d$ states and I $5p$ states compared with lighter halogens. At ambient conditions, $Au^I I_2$ and $Au^{III} I_4$ adopt a linear and square planar configuration respectively (c.f. Figure 1), and have an alternating arrangement, forming a distorted perovskite structure with tetragonal space group I4/mmm [10,11]. The compound undergoes a first order structural transition at 5.5 GPa. Initial experiments indicated that both I(1) and I(2) iodine ions moved to the midpoint of the two Au cations in the horizontal and vertical directions respectively, yielding a tetragonal structure with two equivalent Au sites[5,6,12]. $^{197}$Au Mössbauer spectroscopy measurements conducted on $CsAuI_3$ at 4.2 K suggests the fully $Au^{II}$ SV state occurs at much higher pressures between 6.6 and 12.5 GPa [8]. Applied pressure also strongly affects transport properties. With increasing pressure, the resistivity exhibits at first a reduction in the bandgap, and even a metallic temperature dependence, but the eventual structural transition yields semiconducting behavior, previously attributed to a band Jahn-Teller effect associated with the $Au^{II}$ $5d^9$ electronic configuration [6,12]. A more recent study on the high pressure structure and transport properties of $CsAuI_3$ found that the metallic state was not observed up to 20 GPa, at which pressure the compound became amorphous [13]. Recent pump-probe experiments measuring the life time of the intervalence charge transfer suggest that photo-induced excitation has a similar effect to hydrostatic pressure and further suggests the nonmetallic nature of the excited $Au^{II}$ state[14].

The discrepancies in transition pressure and measured conductivity between previous measurements of $CsAuI_3$ may be due to the resolution of the previous experimental measurements, and the different pressure transmitting media chosen. The alkali metal gold halide family is clearly very sensitive to deviatoric stress: a recent high pressure study on the structure of CsAuCl$_3$ using He as pressure transmitting medium found a pressure-induced transition to a cubic phase at 12.5 GPa [4], contradicting the previous reports that either a cubic structure emerges at 5.2 GPa [15] or a high pressure tetragonal phase appears at 11-12 GPa [5].

In this paper, we present a new study on the high pressure structure of CsAuI$_3$ measured in a hydrostatic environment using synchrotron x-ray powder diffraction. In contrast to previous results, we find that hydrostatic pressure breaks the rotational symmetry of the *ab* plane and a tetragonal to orthorhombic phase transition occurs at 5.5-6 GPa. Our results do not permit a unique determination of the orthorhombic structure, but do clearly point to two possible solutions. We discuss these in the context of the Au valence transition.

## II. EXPERIMENTAL

CsAuI$_3$ crystals were grown by a self-flux method as described elsewhere [16] using CsI, Au and I$_2$ as starting materials. The sample was ground into a powder and loaded into a 150μm diameter sample chamber drilled into a stainless steel gasket compressed between two diamonds with 500μm culet diameter inside a symmetric diamond anvil cell (DAC). To maintain hydrostatic conditions, Ne was chosen as the pressure transmitting medium, which was loaded into the DAC using the gas loading system at GSECARS of the Advanced Photon Source (APS), Argonne National Laboratory (ANL). A small ruby chip placed in the DAC was used for pressure calibration by observing the ruby R1 fluorescence line [17]. Good hydrostaticity was maintained up to the highest pressures measured. *In-situ* high pressure angle-dispersive powder x-ray diffraction (XRD) was conducted at Beamline 12.2.2 of the Advanced Light Source (ALS) at Lawrence Berkeley National Laboratory (LBNL) with incident x-ray wavelength of 0.4959Å. 2D diffraction patterns from MAR345 image plate were integrated using FIT2D [18]. Rietveld refinement was performed on the powder diffraction pattern using the GSAS-EXPGUI package [19].

**III. RESULTS AND DISCUSSION**

Angle dispersive XRD was conducted under hydrostatic compression up to 21.2 GPa. Selected spectra during the increasing pressure cycle are shown in Figure 2. At 5.9 GPa a new high pressure phase starts to emerge, co-existing with the original tetragonal phase. The transition pressure is consistent with the previous results of 5.5 GPa [5,13], but as described below we have assigned the high pressure structure to be orthorhombic. The orthorhombic phase persists until 14.2 GPa, above which the (112)(200)(020) triplet at $2\theta \approx 7.3°$ is nearly gone, and only two groups of broad peaks are present. This observation is consistent with the XRD spectra at similar pressure in Kusmartseva *et al*'s recent study [13], where they pointed out the compound starts to become amorphous above 12 GPa. As shown in Figure 3, during the decompression cycle, crystalline XRD peaks reemerge at 12.5 GPa, indicating the return of the high pressure crystalline phase. As pressure was further released to 4.5 GPa, the diffraction pattern clearly differs from the orthorhombic phase and resembles the low pressure tetragonal phase. Although the diffraction pattern at the lowest pressure during the decompression cycle is not as well resolved as those during the compression cycle, we can clearly see that the pressure induced amorphization of $CsAuI_3$ up to 21 GPa is reversible. This finding contrasts with the previous result that the amorphous phase can be quenched to ambient pressure after being compressed to 32 GPa in a non-hydrostatic environment [13]. Such differences indicate that structural transformation and phase stability in $CsAuI_3$ is very sensitive to the degree of hydrostaticity.

In our XRD spectra (c.f. Figure 4), we see a triplet emerging from the 112 and 200 doublet above 6 GPa that was not reported in previous literature. The absence of this feature in earlier experiments is likely due to diffraction peak broadening that can occur under deviatoric stress when less hydrostatic pressure transmitting media are used. The triplet feature in our spectra is due to the 200 peak splitting into the 200 and 020 in the new orthorhombic phase. The asymmetric feature at $2\theta \approx 9°$ indicates the $2\theta$ value of 202 and 022 reflections are no longer equal. The splitting of these Bragg reflections suggests that the high pressure structure adopts a lower symmetry space group in which the original rotational symmetry in the *ab* plane is broken. We eliminated the possibility of a monoclinic space group that is isostructural to $RbAuI_3$, because this yields a higher volume than the original tetragonal phase. We also notice the persistence of the 103 and 211 features at $2\theta \approx 8.5°$, in contrast to previous results that the features are not observable after the structural phase transition [5,12].

The pressure dependence of the unit cell volume of $CsAuI_3$ is presented in Figure 5. At 5.9 GPa, the orthorhombic phase shows a volume reduction of 0.7%. In the lower pressure tetragonal phase, our measurements give a smaller volume compared to previous results. This could be due to the pressure environment. In the axial diffraction geometry, the Bragg reflections sampled by the detector are those whose planes are nearly parallel to the compression axis. In a non-hydrostatic environment, uniaxial compression tends to result in planes perpendicular to the compression axis being more compressed, and those parallel to the compression axis being less compressed. Therefore, the larger volume in Kusmartseva *et al*'s work maybe a result of their lack of use of any pressure transmitting medium [13]. The slightly different trend in the volume change for the higher pressure phase above 5.5 GPa is due to the different crystal system assignments.

In previous energy-dispersive and angle-dispersive XRD studies, the high pressure phase above 5.5-6 GPa was identified as tetragonal with possible space group *P4/mmm* [5,6,12,13]. In this assignment, the I⁻ ions are located exactly at the midpoint of the two Au ions, making the two Au sites crystallographically equivalent. This result was evidenced by the observation of the weakening and eventual disappearance of 103 and 211 reflections as pressure increases to approximately 5 GPa [5,12]. In our study, however, the persistence of 103 and 211 reflections in both the tetragonal and orthorhombic phase suggests the I⁻ are not in the mid point between the two Au sites of (0,0,0) and (0,0,1/2). There can be a result of at least two types of distortion to this: i) I(2) deviates from the mid point but remains in mid plane of the above mentioned two Au sites, i.e. I(2) is at ($\delta x$, $\delta y$, 1/4); ii) I(2) stays in line with the two Au sites, but deviates from the mid point, i.e. I(2) is at (0,0, $1/4 + \delta z$). Two space groups, *Ibmm* and *Immm*, that fit the diffraction spectra equally well correspond to the two scenarios described above. The difference between these two space groups is that certain Bragg reflections that are forbidden in *Ibmm* space group are allowed in the lower symmetry *Immm* space group. Among the major Bragg peaks of orthorhombic CsAuI$_3$, at the 103 reflection, *Immm* has an additional 013 reflection with a 2θ value $0.028°$ higher than that of 103 reflection. Such features however cannot be unambiguously resolved in our spectra due to strain broadening and limited instrument resolution.

With *Ibmm* or *Immm*, we obtained almost the same lattice parameter for the orthorhombic phase. Their pressure dependence are plotted in Figure 6(A), together with the lattice parameters for the tetragonal *I4/mmm* low pressure phase. At the structural phase transition at 5.9 GPa, the length of the $c$ axis increases by 1.6%, while $a$ and $b$ shrink by 0.3% and 2% compared with $a$ in the tetragonal phase. If we further plot the difference of $a$ and $b$, i.e. the strain of the orthorhombic cell $2(a-b)/(a+b)$ [20] with pressure, we see the value nearly doubles from 7.6 GPa to 12.4 GPa (inset of Figure 6). For the supercell of the perovskite structure (ABX$_3$), the ratio $c/\sqrt{2}a$ represents the average distortion of the BX$_6$ octahedra along the $c$ axis. A value of 1.0 usually implies perfect symmetric BX$_6$ octahedra in the structure. When the structure distorts to an orthorhombic symmetry, we use $c/\sqrt{a^2+b^2}$ to characterize the octahedral distortion instead. In our study, we observed a linear drop of the $c/\sqrt{2}a$ value before the phase transition (consistent with Kojima *et al*'s work [5]), and a sudden increase of the value from 1.016 to 1.045 at the orthorhombic phase. Above 5.9 GPa, the value further climbs to 1.048 at 7.6 GPa and then slowly decreases with pressure. One can see from the decrease of the $a$ and $b$ values and elongation of $c$: the elevated $c/\sqrt{a^2+b^2}$ value at the orthorhombic phase suggests a higher level of distortion in the AuI$_6$ octahedra and a relatively closer distance between the Au at (0,0,0) and (1/2, 1/2, 0) in the $ab$ plane. We further discuss what these observations mean for the two scenarios below.

*A. Single Valent Au$^{II}$ with space group Ibmm*

In this structure, as shown in the top panel of Figure 7, locally equivalent AuI$_6$ octahedra are elongated and tilted towards the $a$ axis. The I(2) ions sit at ($\delta x$, 0, 1/4), breaking the 4-fold rotational symmetry of the $ab$ plane. If the octahedra are observed from the top of the $ab$ plane, their tilting is around the [110] axis. The AuI$_6$ octahedra are the same for all Au sites, which is consistent with a single Au$^{II}$ valence. The structural transition at ~5.9 GPa may then be understood as previously suggested by Kojima *et al* [5,9] and Kitagawa *et al* [12]: with pressure inducing the electron transfer from Au$^{I}$ to Au$^{III}$ (as evidenced by the reduction of $c/\sqrt{2}a$ ratio) in the lower pressure tetragonal phase, the MV-SV valence transition is realized at ~ 5.9 GPa. In the single valence Au$^{II}$I$_6$ octahedron, the Au$^{II}$ 5$d^9$ electron configuration will have a strong Jahn-teller effect, causing the sudden increase of $c/\sqrt{a^2+b^2}$ value. In addition, due to the volume limit imposed by external pressure, the $c$ axis increase cannot provide enough energy, and the AuI$_6$ octahedra have to tilt to reach a longer Au-I distance in the $5d_{z^2}$ direction.

## B. Mixed valent Au$^I$ + Au$^{III}$ with space group Immm

In this structure, the AuI$_6$ octahdra are still distorted but not tilted. The alternation of elongated and compressed octahedra doesn't completely vanish, and, at least from the perspective of symmetry, still comprises a mixed valence state of Au$^I$ + Au$^{III}$. The new cell geometry (longer $c$ axis and closer distances in $ab$ plane) would naturally promote the in-plane Au$^I$ ($5d_{x^2-y^2}$) and Au$^I$ ($5d_{z^2}$) to Au$^{III}$ ($5d_{x^2-y^2}$) charge transfer, which are the dominant intervalence charge transfer interaction according to optical reflectivity measurement and molecular orbital calculations [6]. Hence, in this scenario, while it is possible that intervalence charge transfer still provides the driving force for the structural transition, nevertheless the MV-SV transition remains incomplete.

The high pressure structural transition of CsAuI$_3$ revealed by our study clearly differs from that of CsAuCl$_3$ and CsAuBr$_3$. The chloride undergoes a first order tetragonal *I4/mmm* Z=2 to cubic *Pm3m* Z=1 structural transition at 12.5 GPa, shown by single crystal x-ray diffraction, and its cubic structure naturally leads to a single Au site, with a formal valence Au$^{II}$ [4]. The bromide was shown to undergo a tetragonal *I4/mmm* Z=2 to tetragonal *P4/mmm* Z=1 transition at approximately 9 GPa [5], at which pressure the two Au sites also became crystallographically identical. The associated MV-SV transition is consistent with the observation from Raman spectroscopy [7], and charge density analysis from synchrotron powder XRD[21]. In the case of CsAuI$_3$, the structural transition observed at ~5.9 GPa that breaks the tetragonal symmetry has never been observed in the chloride and bromide. If concurrent with the MV-SV transition, it is caused by the Au$^{II}$ 5$d^9$ Jahn-Teller effect. If not concurrent with the valence transition, it further promotes the charge transfer between Au$^I$ and Au$^{III}$. Previous Mössbauer study showing that a single Au valence is achieved between 6.6 GPa and 12.5 GPa[8,9] seems to support the notion of an incomplete valence transition at ~5.9 GPa. However, one must note that the measurement was conducted at 4K, at which temperature the first order transition can be very sluggish. Even in our 300 K measurement, we see a co-existence of the two phases at 5.9 GPa, and it is possible that a larger hysteresis occurs at lower temperature.

In summary, we report reversible tetragonal-orthorhombic-amorphous structural transitions in CsAuI$_3$ in a hydrostatic environment up to 21 GPa with critical pressures of 5.5-6 GPa and 12-14 GPa respectively. While our results may suggest the MV-SV electronic transition is driving the structure transition (as in the case of *Ibmm* space group), we cannot rule out the possibility that the valence transition is not completed at the structural transition (if transit to a structure of *Immm* space group), and the charge disproportionation in Au persists into the high pressure orthorhombic phase. The observation of the orthorhombic phase calls for a more careful study of the structural, electronic and magnetic properties of CsAuI$_3$ at high pressure to reassess the pressure-induced charge transfer, and the MV-SV transition in CsAuI$_3$.

**Acknowledgements**

We like to thank our reviewer for pointing out an important aspect of our data. SW, SH, and WLM are supported by EFree, an Energy Frontier Research Center funded by the U.S.

Department of Energy (DOE), Office of Science, Office of Basic Energy Sciences(BES) under Award Number DE-SG0001057. MCS, SCR, THG and IRF are supported by the Airforce Office of Scientific Research (AFOSR) under grant FA9550-09-1-0583. Travel to experimental facilities are supported through DOE-NNSA(CDAC). GSECARS is supported through COMPRES under NSF Cooperative Agreement EAR 10-43050. ALS is supported by DOE-BES under Contract No. DE-AC02-05CH11231.

**Figures**

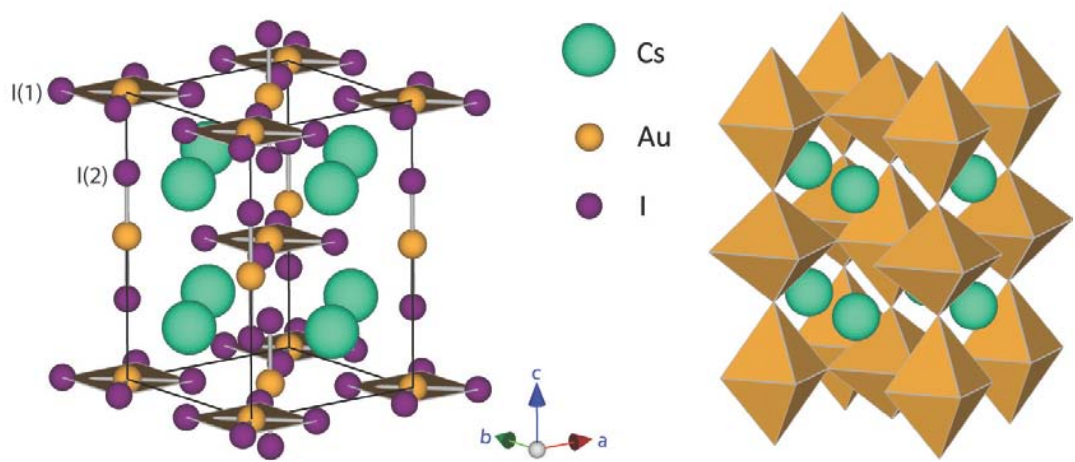

**Figure 1**. Ambient pressure crystal structure of CsAuI$_3$ at 300 K. Left panel shows the linear and square planar coordination of Au$^I$I$_2$ and Au$^{III}$I$_4$ respectively. We denote the I as I(1) for those associated with square planar Au$^{III}$I$_4$, and I as I(2) for those with the linear Au$^I$I$_2$. Right panel shows the structure in alternating elongated and compressed AuI$_6$ octahedra, more clearly revealing the distorted perovskite structure.

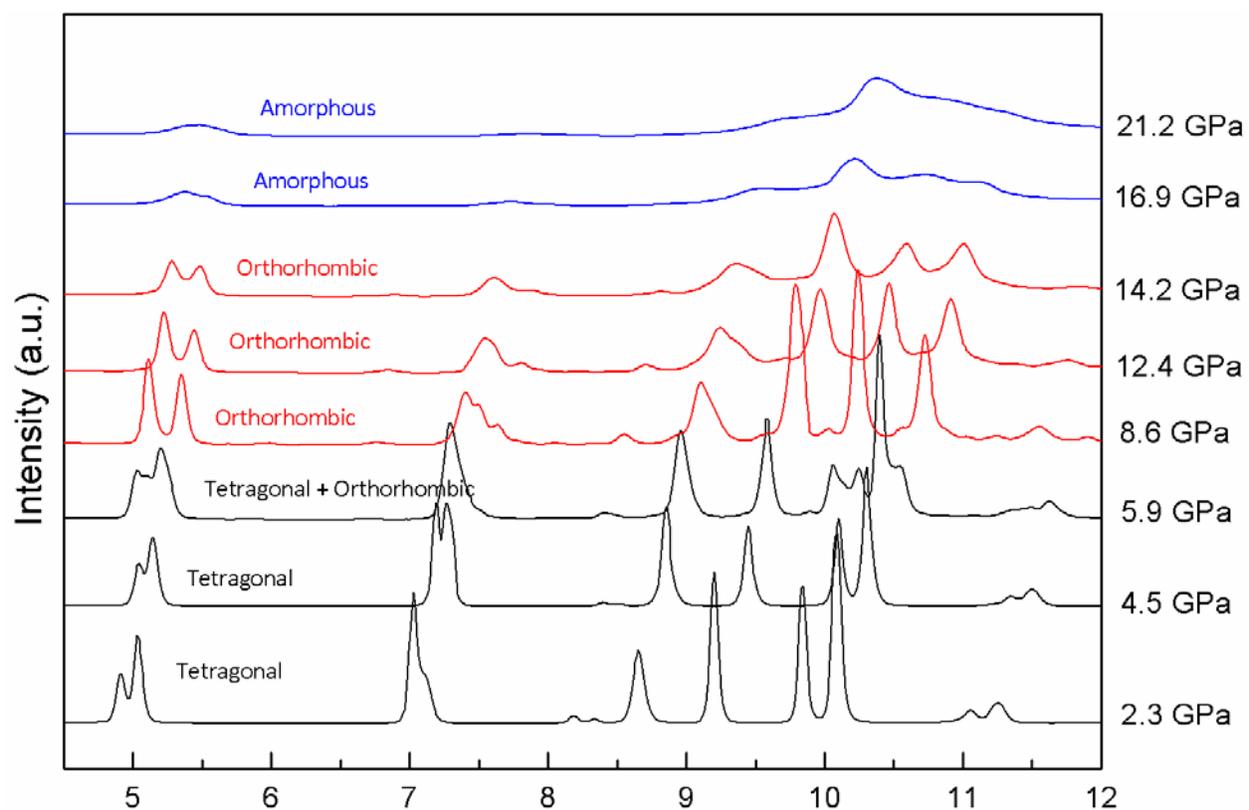

**Figure 2**. Selected angle dispersive x-ray powder diffraction patterns for CsAuI$_3$ at 300 K taken upon increasing pressure. Black curves: tetragonal phase, red curves: orthorhombic phase and blue curves: amorphous phase. At 5.9 GPa there is a co-existence of the tetragonal and orthorhombic phases.

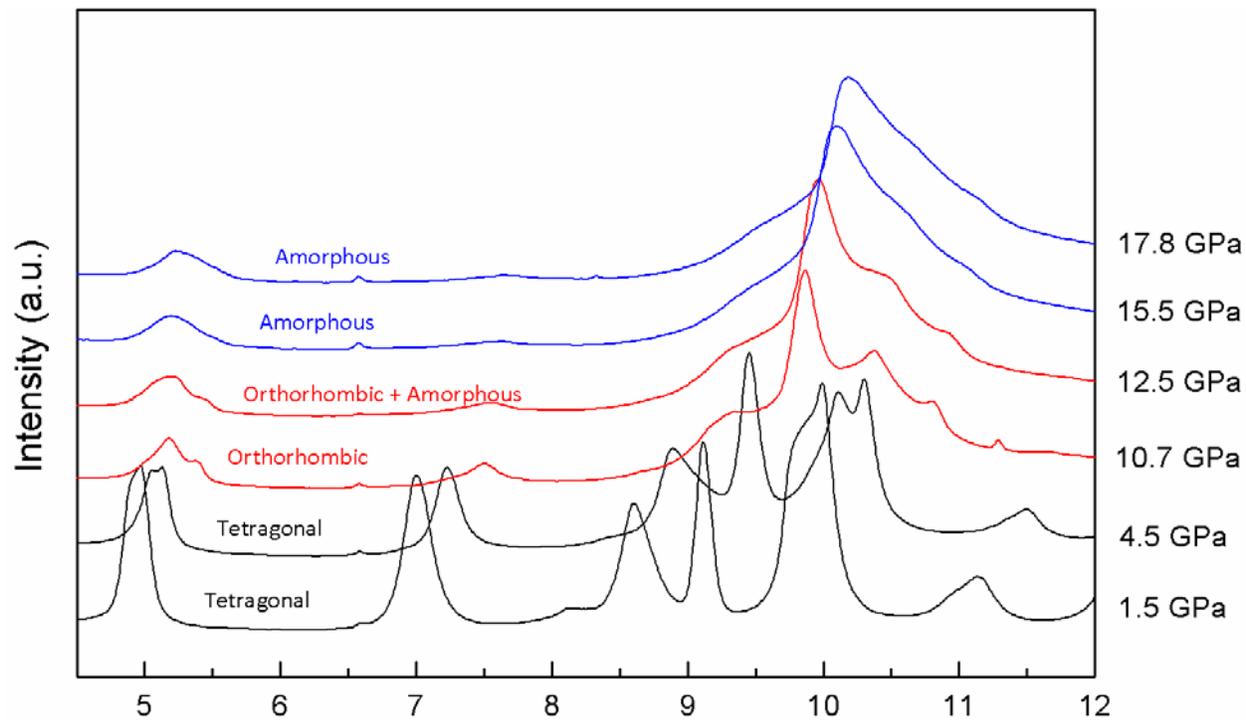

**Figure 3**. Selected angle dispersive x-ray powder diffraction patterns for CsAuI$_3$ at 300 K taken upon decompression cycle. Black curves: tetragonal phase, red curves: orthorhombic phase and blue curves: amorphous phase. The patterns indicate that the amorphization is reversible.

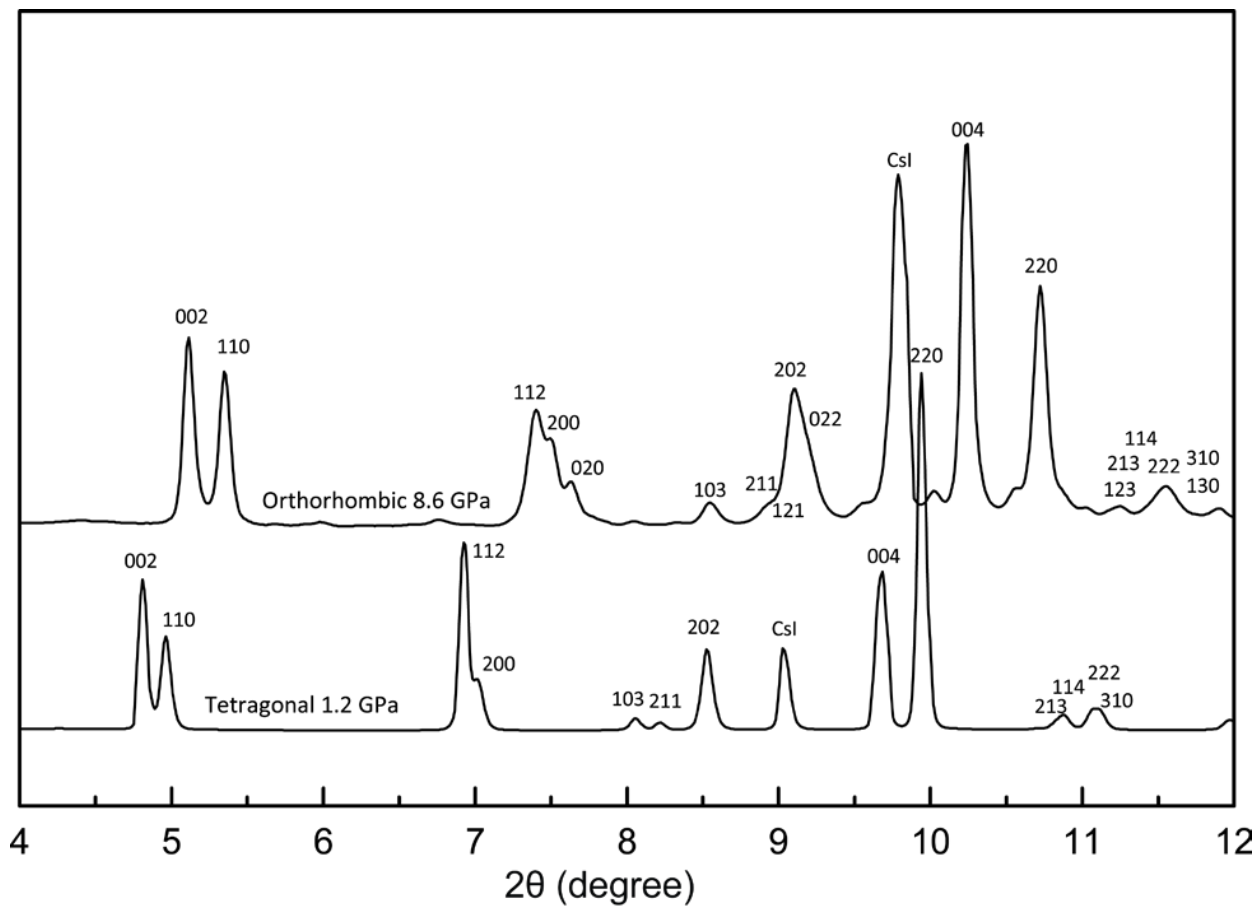

**Figure 4**. Indexed XRD patterns for the low pressure tetragonal phase at 1.2 GPa and high pressure orthorhombic phase at 8.6 GPa. The 112 and 200 doublet feature in the tetragonal phase evolves to a triplet of 112, 200 and 020 reflections in orthorhombic phase. The peak marked with CsI is the strongest 110 reflection of simple cubic CsI.

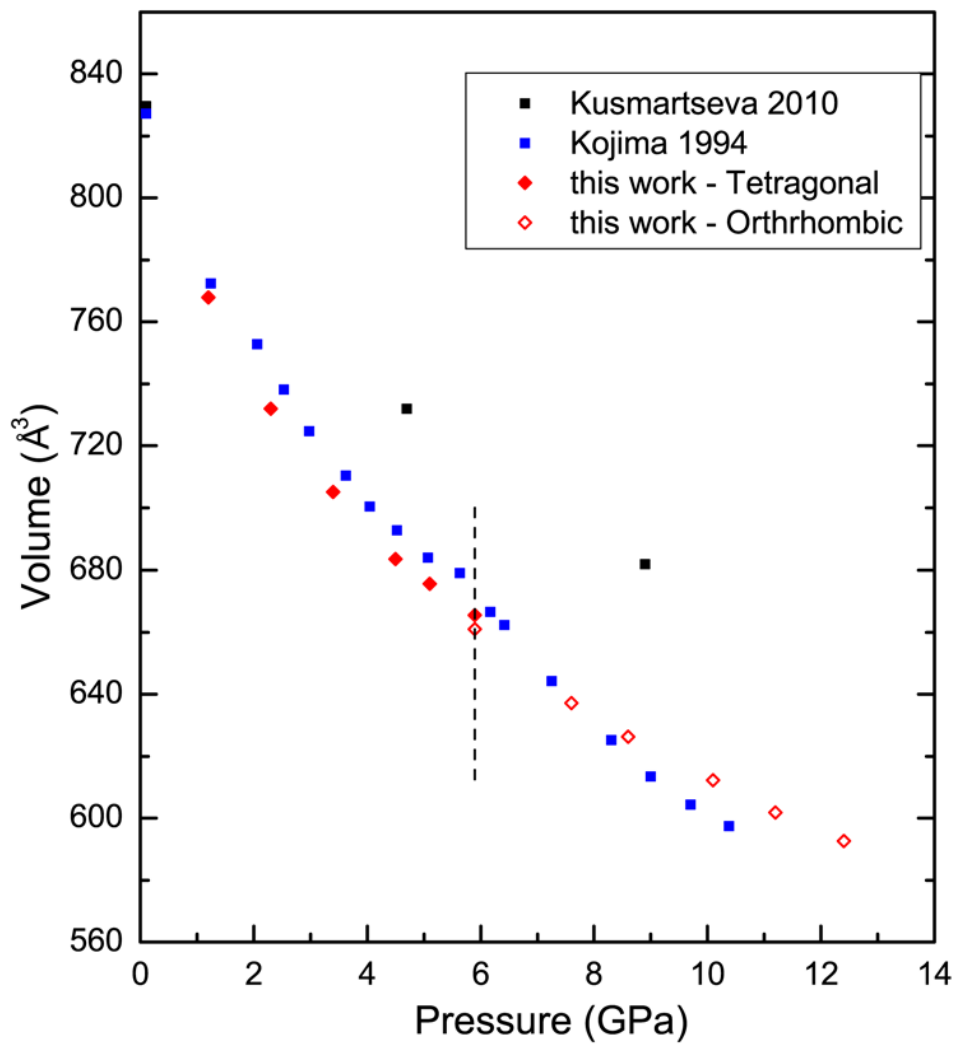

**Figure 5**. Evolution of the unit cell volume of $CsAuI_3$ with pressure at 300 K compared with previously published results. Black squares represent data from Ref. 12, blue squares are from Ref. 5. Red diamonds are from this work: unfilled symbols represent low pressure tetragonal phase and filled symbols represent high pressure orthorhombic phase. Vertical dashed line marks onset of 1st-order pressure-induced phase transition for increasing pressures.

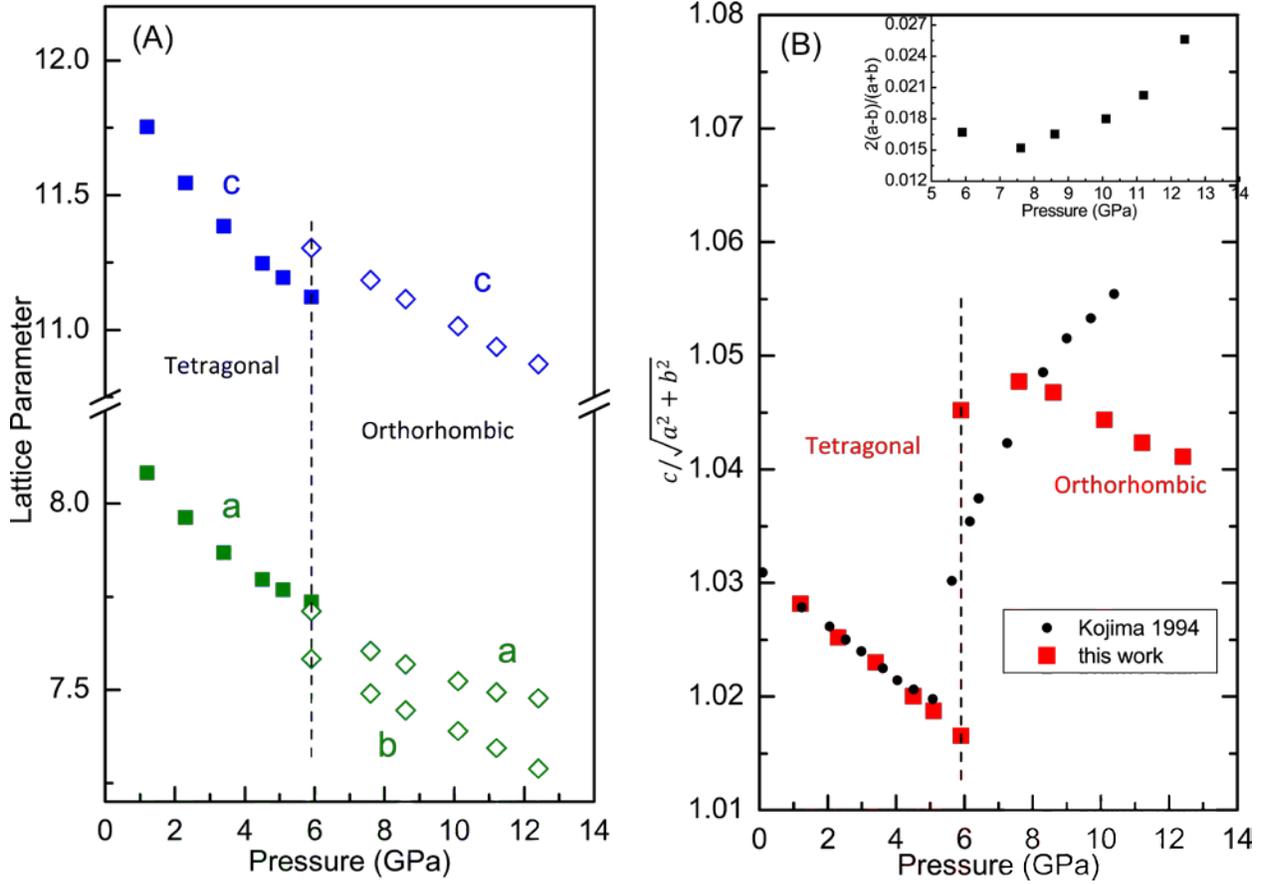

**Figure 6**. (A) Lattice parameters of $CsAuI_3$ as a function of pressure. Green symbols: $a$ in tetragonal and $a,b$ in orthorhombic phases, blue symbols: $c$. (B) $c/\sqrt{2}a$ and $c/\sqrt{a^2+b^2}$ for tetragonal and orthorhombic phases. Black dots are from previous work which fitted all the phases with tetragonal space group [5]. Red squares represent this work. Inset: $2(a-b)/(a+b)$ describing the strain of orthorhombic cell as a function of pressure. Vertical dashed line marks onset of first-order phase transition for increasing pressures.

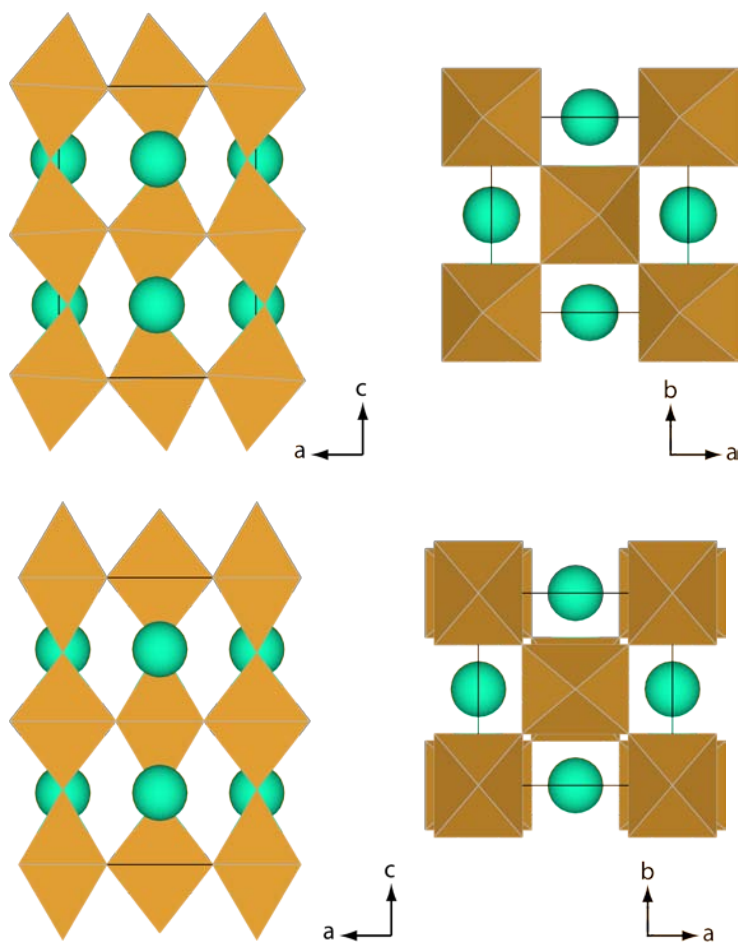

Figure 7. Schematic drawing of the AuI$_6$ octahedra projection in orthorhombic CsAuI$_3$ at 8.6 GPa. Top panel: Structure with space group *Ibmm*: locally equivalent octahedra rotate toward $a$ axis, and around [110] plane. Bottom panel: Structure with space group *Immm*: elongated and compressed octahedra arrange alternatively, and do not tilt.